# Potential Tribological and Antibacterial Benefits of Pulsed Laser Deposited Zirconia Thin Film on Ti6Al4V Bio-alloy

*S. Kedia[1,3], A. Das[2], B. S. Patro[2], J. P. Nilaya[1,3]

[1]Laser & Plasma Technology Division, Bhabha Atomic Research Centre, Mumbai 400085

[2]Bio-Organic Division, Bhabha Atomic Research Centre, Mumbai 400085

[3]Homi Bhabha National Institute, Training School Complex, Anushaktinagar, Mumbai 400094

*Corresponding author: skedia@barc.gov.in

**Abstract:** Demand for artificial body implants has been on the rise over the years. However, wear and bacterial infection are identified as two major reasons that can lead to inflammation and implant failure. In this communication, the advantages of pulsed laser deposited zirconia thin film on Ti6Al4V bio-alloy at room temperature and at an elevated substrate temperature are discussed wherein a comparison of the change in surface roughness, wettability, surface free energy, tribological and antibacterial properties of uncoated and zirconia coated Ti6Al4V samples is presented. The results of tribological analysis carried out using a standard ball-on-disc tribometer at different loads (2N, 5N and 7N) exhibited advantageous effects of zirconia coating on Ti6Al4V. Prominently, the sample coated at 200 C substrate temperature maintained very low coefficient of friction up to hundreds of sliding cycles and showed a notable reduction in the wear rate by 49% at 5N load. The *in vitro* bacterial retention test showed a clear inhibition in growth of *Staphylococcus aureus* and *Klebsiella pneumonia* bacteria on the surface of the coated samples indicating the possibility of prevention of biofilm formation. More than 50% reduction in density of *Staphylococcus aureus* was observed on coated sample in comparison to pristine Ti6Al4V and this can be attributed to reduction in surface energy of the sample after coating. Additionally, the observation of a larger number of decimated bacteria on coated samples by fluorescence microscopy revealed superior antibacterial properties of zirconia coating. The novelty of this work is the use of pulsed laser deposition technique for zirconia coating which dearly improves tribological and antibacterial properties of Ti6Al4V simultaneously; this shows prospects of increasing durability of artificial implant in the human body.

1. **Introduction**

As the demand for artificial implants in orthopedic fixation, dentistry, stents, and tissue engineering is increasing, research activity towards improving their quality and longevity has garnered attention [1, 2]. Based on important properties such as biocompatibility, mechanical stability, and non-toxicity some materials e.g., titanium alloys, stainless steel, cobalt alloys, aluminum oxide, zirconia, calcium phosphates, and poly(methyl)metahcrylate have been recognized as suitable biomaterials to construct artificial body implants [3]. However, the issue of early and late implant failures in patients caused by poor osseointegration, corrosion, wear, or bacterial infections remains a cause of concern [4]. A satisfactory osseointgration of the implant in the body ensures its primary stability and eliminates the risk of fibrous tissue formation at the bone-implant interface, both these factors being responsible, more often, for an implant failure. An implant faces unavoidable wear and corrosion in the harsh body environment. A biomaterial with poor corrosion resistance can degrade and release toxic ions in the body leading to complications. Moreover, an implant with a high wear rate can generate wear debris which can expedite dimensional instabilities [5]. This debris can be in the form of free metallic ions, colloidal complexes, metal oxide, or metal particles, all of which have a very larger surface area to interact with surrounding tissues. Under certain circumstances, this debris can induce inflammatory reactions and allergies in the patient. This complex process of wear is largely dependent upon the contact area between the two surfaces and the sliding distance, both of which increase gradually with physical activity and weight gain of the patient [6]. Another major issue that is accountable for implant failure is bacterial infection and biofilm formation on the implant surface [7]. Bacterial adhesion can, in some cases, lead to bacterial colonization and biofilm formation as well. The general sources of bacteria include the environment of the operation room/hospital, surgical equipment, bacteria already present on the patient's skin or body. Bacterial infection during or after implantation can lead to early complications or permanent contamination and remedy for both these situations accounts for an enormous medical cost, the surgical removal of the infected implant becoming the sole solution in some cases [8]. All these factors lead to a decrease in the service life of an artificial implant and therefore, the need for the development of biomaterials with superior properties.

Among mentioned biomaterials, grade-5 titanium alloy (Ti6Al4V) has been considered as the most promising material for orthopaedic due to its inherent properties of low density, high tensile strength, excellent biocompatibility and good corrosion resistance [9] and, the near similar values of elastic modulus of Ti6Al4V (55-110 GPa) and natural bone (10-40 GPa) [10]. However, Ti6Al4V has a high coefficient of friction (CoF) and low wear resistance which restricts its uses in load-bearing applications [5, 11]. Alteration in the surface properties either by topographical modification [12] or by suitable surface coating [13] is an alternative route to achieve desired functionalities in a bio-material. Various materials such as, diamond-like carbon, graphite-like carbon, tantalum and titanium nitride, hydroxyapatite, zirconia, alumina have been used as coating materials for biomaterials [14-16] by sol-gel [17], pulsed laser deposition (PLD) [18], chemical vapor deposition [19], or RF magnetron sputtering [20] techniques. PLD, a technique capable of providing a precise control over thickness of deposition, maintaining stoichiometry of the target material, freedom of using versatile targets, and option of increasing substrate temperature during deposition, is a sought after method to modify the surface of a biomaterial [13]. A surface coating that can simultaneously improve wear resistance and inhibit bacterial adhesion on Ti6Al4V surface without losing its bulk mechanical properties will ensure a better tissue integration and also stretch the service life of an implant. A promising candidate meeting these requirements is zirconia ($ZrO_2$) which is a crystalline dioxide of zirconium, and is a widely accepted biomaterial in dental and orthopedic because of its elevated hardness, good biocompatibility, and antibacterial properties [21-23]. The direct use of $ZrO_2$ as an implant can be costly, however, its characteristics can be imparted to the implant coating its thin film on Ti6Al4V.

This work is aimed to assess the potential tribological and antibacterial benefits of $ZrO_2$ thin film coating by PLD on Ti6Al4V bio-alloy. The PLD was performed with the substrate being maintained at room temperature and at an elevated temperature ~200 C. The third harmonic of a nanosecond Nd:YAG laser (Model # N311, Ekspla make) capable of delivering a max of ~ 150 mJ/pulse at 355 nm at a repetition rate of 10 Hz was employed as the ablation source. The improvement in the biofunctionality as a result of $ZrO_2$ coating and the effect of substrate temperature on this process were studied. The surface morphology, roughness, adhesion, wettability, and surface free energy of the samples were recorded using a scanning electron microscope (SEM), Atomic Force Microscope (AFM), scratch test, static sessile drop technique, and geometric-mean methods, respectively. The tribological analysis was carried out in ambient using standard ball-on-disc method at 3 different loads of 2N, 5N and 7N. A clear reduction in CoF of ~ 23% at 2N and wear rate up to 49% at 5N were observed for sample coated at 200 C substrate temperature. To our knowledge, such extent of tribological behaviour of PLD coated $ZrO_2$ on Ti6Al4V has not been reported. The antibacterial properties of samples were tested against two bacteria; gram positive Staphlococcus aureus (*S. aureus*) and gram negative Klebsiella pneumoniae (*K. Pneumoniae*). Adhesion and growth of both the bacteria were found to be reduced on coated samples as studied by total viable count. A qualitative confirmation was obtained using confocal microscopy where defeat in bacterial colony, biofilm formation and the more percentage of decimated bacteria was found on the coated samples. A method of improving both, the wear resistance and antimicrobial properties of Ti6Al4V by PLD of $ZrO_2$ is demonstrated.

2. **Materials and methods:**

   *2.1 Materials:*

As received $ZrO_2$ pellet (M/s. Merck, KGaA) of thickness 2 mm and diameter 10 mm were used as PLD targets. Ti6Al4V substrate of dimensions (20 mm x 20 mm x 1 mm) were surface polished using SiC papers of grit size 400, 800, and 1200. The polished samples were cleaned with acetone, ethanol, and water for 10 min each followed by ultrasonic cleaning and stored in a desiccator till further use.

   *2.2 PLD of ZrO2 thin film:*

PLD is a physical process in which a focused pulsed laser beam ablates the desired target material and a thin film deposition occurs on an appropriately positioned substrate in a high vacuum background (~2 x $10^{-5}$ mbar). Fig. 1 illustrates the experimental set up of PLD chamber used for deposition of $ZrO_2$ on Ti6Al4V. The laser beam (λ = 355nm, pulse duration

6 ns, repetition rate 10 Hz) was focused using a lens ($f = 50$ cm) onto a ZrO$_2$ pellet placed at a distance of 40mm from the substrate. A deposition time of 1 hour was found to be optimum to obtain a uniform layer when the laser fluence was maintained at 20 J/cm$^2$. Provision was made to change the temperature of the substrate using a heater and film deposition was carried out at two substrate temperatures viz., room temperature and 200 C and the corresponding samples are referred to as S-RT and S-200, respectively. The results of the ZrO$_2$ coated samples were compared with the results of the pristine sample termed as S-P. Since deposition occurs in a clean environment, the transfer of target to the substrate is stoichiometric, and the thickness of the film can be controlled by the number of pulses and duration of the deposition process.

### *2.3 Surface characterization:*

#### *2.3.1 Topography, film thickness, and surface roughness*

Topography and surface roughness of samples before and after ZrO$_2$ coating were analyzed from SEM and AFM images, respectively. Adhesion of ZrO$_2$ thin film on Ti6Al4V surface was analyzed by Rockwell scratch test using a diamond indenter of radius 100 μm (M/s. Anton Paar). Progressive load starting from 0.03 N to 5N with the rate of 4.97 N/min was applied during the measurement and speed was maintained at 0.5 mm/min for a length of 0.5 mm.

#### *2.3.2 Wettability and Surface free energy*

The interaction between the surface of a biomaterial and its environment, e.g. bacteria in an aqueous solution depends upon the chemistry of both and can be estimated from their interfacial energy that is a direct measure of the intermolecular attractive forces. The adhesion of bacteria on the sample surface is governed both, by dispersive as well as polar interactions and hence the interfacial energy comprises of the Lifshitz-van der Waals attractive interaction component, the electrostatic double-layer repulsive component, the Lewis acid-base component, and the Brownian motion component. Microbial adherence to the sample surfaces occurs when the total interaction energy is negative [24]. Direct measurement of interfacial surface energy is difficult; however, it can be estimated from a number of contact angle (CA) measurements of polar and non-polar liquids of known surface tension on the sample surface using standard software.

In the present work, two polar solvents; water and formamide and a non-polar solvent; xylene with known surface tension have been used. The change in surface energy of uncoated and $ZrO_2$ coated Ti6Al4V samples was directly obtained from the software by feeding contact angle data of all three liquids. The CA were evaluated using static sessile drop technique with M/s. VCA optima analyser. A software controlled micro-syringe was used to drop liquid of volume ~ 1 μL on the sample and photo of the drop was recorded immediately using an inbuilt CCD camera. The drop image was processed using image processing software to record advancing ($\theta_a$) and receding ($\theta_r$) CA from the shape of the drop. Three measurements with each liquid at three different locations on the sample surface were performed to determine average CA of all samples.

### 2.3.3 Tribological tests:

The dry sliding wear tests were carried out by a ball-on-disc tribometer (M/s. Anton Paar). A standard stainless steel (SS) ball of diameter about 5 mm was used as counterpart. Sliding friction and wear tests were measured at constant normal load of 2N, 5N and 7N at room temperature in air. During the wear tests, the sample was fixed and SS ball was rotated against the stationary sample at liner velocity of 300 mm/min for 2000 numbers of cycles and variation in CoF was directly recorded during the rotation. The wear area on the sample was estimated using a stylus profilometer and wear area on counterpart was obtained from optical microscope image. The wear rate was directly calculated by the software. The wear mechanism was studied by topographical analysis of wear traces in SEM.

### *2.3.4 Bacterial adhesion tests:*

*S. aureus* MTCC 7373 and *K. pnemoniae* MTCC 109, were isolated in Nutrient Agar and grown in Nutrient broth. The bacteria were grown overnight in Nutrient Broth followed by centrifuging. The supernatants were discarded and pellets were re-suspended in phosphate-buffered saline (PBS), followed by a second centrifuging and resuspension in PBS. The OD value of the cell suspension was measured using a spectrophotometer at an excitation wavelength of 600 nm, and the value was adjusted to 0.1 by gradually adding PBS [25]. The samples were placed separately in 60 mm × 15 mm sterilized Petri dishes followed by the addition of 500 µL of the bacterial suspension. The samples were incubated at room temperature with intermittent mixing by pipette for 4 h. At the end of the incubation period, the suspension was gently pipetted from the surface and the plates were then put into a new tube containing 5 ml of PBS, and vigorously vortexed for 30 s to remove the adhering micro-organisms. The pipetted culture was mixed with the vortexed culture, centrifuged, and re-suspended in PBS. Viable organisms were quantified by plating serial dilutions on Nutrient agar plates. Nutrient agar plates were incubated for 24 h at 37°C and the colony-forming units (CFU) were counted visually.

For the staining procedure, two fluorescent dyes were used in combination: SYTO9 (Invitrogen AG, Basel, Switzerland), and PI (Invitrogen). Stock solutions of the dyes were prepared as follows: PI and SYTO9 were used from the LIVE/DEAD BacLight system as proposed by the manufacturer. All stock solutions were stored at 20°C. All the samples, S-P, S-RT and S-200 were cultured ($10^6$ cells/ml) immediately stained with a mixture of SYTO9 (5 µM final concentration) and PI (30 µM) [26]. Samples were incubated in the dark at room temperature for 15 minutes before analysis. The samples were washed once in PBS and 2 µL was spotted on 1% agarose-coated slides. Fluorescence microscopy analysis of the slides was performed using an LSM 780 Meta laser scanning confocal microscope (Carl Zeiss, Oberkochen, Germany) with a 60 × 1.4 NA Plan-Apochromat oil immersion objective [27].

## 3. Results

### 3.1 Surface analysis

Fig. 2 shows, SEM images, AFM images, scratch test results, and water contact angle of S-P, S-RT, and S-200 samples. Change in the morphology of the surface after deposition of $ZrO_2$ thin film was observed in the SEM images. The density of $ZrO_2$ globules was more on S-RT. However, the average size of $ZrO_2$ particles was larger on S-200. AFM was used to measure the change in average surface roughness of the sample. A reduction in surface roughness from 218 nm of S-P to 104 nm and 82 nm for S-RT and S-200 respectively was observed. The optical images of scratch lines are shown in Fig. 2 where an increase in scratch width with the progressive load was observed on S-P. Complete delamination of $ZrO_2$ film was observed on S-RT at 0.4 mm distance (marked with a circle), however, the film was intact in S-200 at the same distance. This indicates superior mechanical property and better adhesion of $ZrO_2$ film in the case of S-200. The photograph of the water drop on S-P revealed its hydrophilic nature with a water CA of 78°. The wettability of the sample decreased and the sample became hydrophobic with CA of 100° (S-RT) and 103° (S-200) after $ZrO_2$ coating. To know the change in surface energy of Ti6Al4V after $ZrO_2$ coating, the advancing and receding CA of polar solvents water, formamide, and non-polar solvent xylene were measured on the left and right sides of a 2D image of 3 drops at three different locations. Measured values of the CA are listed in table-I. While the CA for formamide increased on coated samples, it showed a reverse trend for xylene for which the CA decreased as a result of coating. The resulting surface free energy of all samples was calculated using these CA values and mentioned in table-I. The surface free energy of Ti6Al4V can be seen to reduce from 18.5 dyne/cm to 17.5 dyne/cm for S-RT. This value further reduced to 14.9 dyne/cm for S-200.

**Table-I:** *Advancing ($\theta_a$) and receding ($\theta_r$) contact angles of different liquids on S-P, S-RT, and S-200 samples*

| Sample/ Solution | WATER | | FORMAMIDE | | XYLENE | | Surface Energy (dyne/cm) |
|---|---|---|---|---|---|---|---|
| | $\theta_a$ | $\theta_r$ | $\theta_a$ | $\theta_r$ | $\theta_a$ | $\theta_r$ | |
| S-P | 75.60 | 74.60 | 53.30 | 57.90 | 16.40 | 20.70 | 18.5 |
| | 91.30 | 91.30 | 60.80 | 61.20 | 24.50 | 21.70 | |
| | 79.30 | 80.60 | 53.90 | 53.80 | 14.90 | 18.70 | |
| S-RT | 95.40 | 96.10 | 84.80 | 85.20 | 12.30 | 13.00 | 17.5 |
| | 100.60 | 100.50 | 85.50 | 86.00 | 15.00 | 14.00 | |
| | 103.70 | 102.30 | 85.50 | 85.50 | 11.80 | 10.80 | |
| S-200 | 101.70 | 101.90 | 99.50 | 99.00 | 5.40 | 5.20 | 14.9 |
| | 103.20 | 104.20 | 89.10 | 91.10 | 3.70 | 3.60 | |
| | 103.20 | 103.80 | 80.50 | 78.80 | 7.80 | 3.50 | |

*3.2 Tribological behaviour*

Table-II: *Minima and maxima of CoF for S-P, S-RT and S-200 samples*

| Samples / CoF | 2N | | 5N | | 7N | |
|---|---|---|---|---|---|---|
| | Min | Max | Min | Max | Min | Max |
| S-P | 0.63 | 0.92 | 0.45 | 0.56 | 0.42 | 0.45 |
| S-RT | 0.27 | 0.92 | 0.18 | 0.56 | 0.16 | 0.56 |
| S-200 | 0.19 | 0.63 | 0.13 | 0.60 | 0.13 | 0.43 |

Fig. 3a-3c shows variation in CoF of S-P, S-RT, and S-200 samples as a function of sliding cycles at different loads of 2N, 5N, and 7N, respectively. The CoF for S-P exhibited a sharp rise to a value of 0.63 during the initial few cycles and with an oscillating behavior reached 0.92 at the end of 2000 cycles. In case of S-RT, the minimum value of CoF at 0.28 is sustained till ~100 cycles after which it suddenly rose to 0.59. This clearly points to the fact that after 100 cycles the counterpart pierced the ZrO2 coating and touched the Ti6Al4V surface. A significant reduction in CoF was observed for S-200 where, the initial CoF was only 0.19, that is about 23% lesser than S-P and that sustained up to ~1500 cycles, following which the CoF rose to a maximum value of 0.63. The similar trend of improvement in CoF in case of S-200 as compared to S-RT and S-P was observed at higher loads too although the break point (the point where there is a sharp rise in the CoF) occurred for lower number of cycles. For a load of 5N, the minimum CoF for S-RT and S-200 remained intact (about 30% lower value for both) up to ~143 and ~772 cycles, respectively after which both samples showed a sharp rise in CoF and reached the same value as S-P. At a higher load of 7N, S-P showed almost constant CoF from starting while S-RT and S-200 had lower CoF up to ~90 and ~327 cycles, respectively which increased at the end and was highest for S-RT. The exact minimum and maximum values of CoF of all samples under all load conditions are listed in table-II for comparison.

To calculate wear rate, trace area was estimated from the resultant wear track on the sample using a stylus profilometer and the depth profile of all samples is shown in Fig. 4a where the wear depth indicates the amount of material removed from the sample, and trace width represents the contact area between the sample and the counterpart at the end of wear cycles. In general, S-P sample exhibited a deeper wear groove compared to S-RT and S-200 samples. The wear depth increased with increased loads. A significant reduction in the wear rates of 24%, 49% and 46% were observed at 2N, 5N and 7N, respectively for S-200 sample in comparison to S-P, as shown in Fig. 4b.

Fig.5 shows the SEM images of wear traces of samples after 2000 cycles at load of 2N, 5N, and 7N values, respectively. Continuous wear tracks with groove structures therein were commonly observed in all cases, indicated occurrence of abrasive wear on the samples. At a load of 2N, tearing of the film and accumulation of wear debris at some location in the trace (marked with a circle), was observed, this clearly indicated adhesive wear. At a load of 5N, along with debris, craters, marked in white rectangles in the figure, were observed in the wear track. At the highest applied load of 7N, extensive wear occurred on the sample surface, and deformations were found on the surfaces, marked with arrows in the figure.

Since the wear resistance of S-200 was superior, EDS analysis of the worn surface was done for this sample, and the results are shown in Fig. 6. The presence of Fe and O in wear trace indicated deposition of debris from SS counterpart on sample after adhesive wear. Furthermore, this result indicated the occurrence of oxidization wear on the sample. The wt% of Fe and O reduced at higher loads while wt% of Ti increased indicating the complete wearing off of the coating at higher load.

### *3.3. Antibacterial properties*

The result of the total viable count of *S. aureus* and *K. pneumonia* after exposure to S-P, S-RT and S-200 is shown in Fig. 7a and 7b respectively, where higher inhibition of both strains on S-RT and S-200 samples compared to control and S-P was clearly observed. These experiments were repeated 4 times and the statistical parameters were calculated. The peptidoglycan rich gram positive *S. aureus* was most resistant by S-RT while showed relatively lesser inhibitory effect by S-200 (Fig. 7a). It is worth noting that the gram negative *K. pneumonia* was equality sensitive to all the treated surfaces with slightly more inhibition on S-200 (Fig. 7b).

To verify our findings at the cellular level, we stained the treated bacteria with LIVE/DEAD BacLight system. Living cells with intact membrane allow only permeable dyes like SYTO9 and exclude membrane impermeable dye like PI, while injured/dead bacteria with compromised membranes allow PI to permeate the cell. Thus living bacteria stain green, while injured/dead cells stain red. It can be clearly observed that, bacterial adhesion, colony and biofilm formation are more on S-P and S-RT samples for both microbial. However, lesser bacteria were adhered on S-200 and the number was even lesser in case of *S. aureus*. We found that, for both *S. aureus* (Fig. 8a) and *K. Pneumonia* (Fig. 8b), the proportion of red stained cells increased after exposure to the coated Ti6Al4V samples, thus indicating that these cells have suffered substantial damage to the cell wall/outer membrane. Thus there was an agreement of our cytological data with the physiological data of the viable count, which confirms the bactericidal properties of the treated surfaces. This indicated $ZrO_2$ coating inhibited the adhesion of bacteria and decimated a larger fraction of it within 4hrs.

4. **Discussion**

Ti6Al4V alloy has been used for the production of orthopaedic and dental implants since decades. However, efforts are being expended to enhance its functionalities and thereby increase the overall lifespan of the implant. We have experimentally shown that PLD of $ZrO_2$ thin film on Ti6Al4V surface improves its wear resistance and reduces the interfacial surface energy that in turn, results in a superior antibacterial performance. Results are discussed in detail in the following paragraphs.

*4.1 Tribological behaviour*

The value of CoF is an indicator of the tribological behaviour of the material where smaller CoF corresponds to superior wear resistance [28]. Section 3.1 reported results of CoF and wear rate of S-P, S-RT and S-200 samples characterized using ball-on-disk tribometer under dry sliding condition against SS counterpart ball at different loads. For all loads, after initial few cycles, a sharp rise in the CoF for S-P was observed which remained high throughout the test consisting of 2000 cycles. The initial lower value of CoF for S-P is because of thin oxide layer present on its surface that tears of within a few cycles and the counterpart touches Ti6Al4V. The $ZrO_2$ coating seems to have effectively reduced the CoF of Ti6Al4V, more prominently so for S-200. Further, the analysis on S-RT indicated that, the strength and adhesion of $ZrO_2$ coating performed at room temperature could prevent the penetration of the counterpart for some more number of cycles before tearing off. It may be noted that during the initial cycles, the small contact area between the sample and counterpart results in a higher pressure. As soon as counterpart penetrates the surface, the contact area increases and a steep increase in CoF occurs. The non-uniform nature of penetration is reflected in the oscillatory behaviour of CoF the average value of which gradually increases with an increasing number of cycles [29]. That S-200 could withstand with much lower CoF value for more number of cycles at all loads before giving way can be attributed to the superior quality of thin film deposited at higher substrate temperature. These observations are further corroborated by the results of the scratch test shown in Fig. 2 too where delamination occurs at a much later distance for S-200 as compared to others under identical conditions. $ZrO_2$ (653 HV) is known to possess greater hardness as compared to Ti6Al4V (357 HV) [30]. Our experiments reveal that the coating at an elevated temperature (200 C in the present case) plays a decisive role in the improving quality of adhesion. It was observed that the number of cycles up to which $ZrO_2$ sustain for S-200 reduced with increasing load. The wear depth measured using a mechanical profilometer (Fig. 4a) revealed a reduction in wear depth in coated samples as the hardness of $ZrO_2$ restricted penetration of counterpart. About 50% reduction in wear rate indicated superior protectiveness of PLD coated $ZrO_2$ film at 200 C substrate temperature. The SEM images (Fig. 5) showed typical morphologies of the worn surface at two magnifications (300 μm and 50 μm). Continuous sliding marks with grooves and ridges seen in all wear tracks resulted from penetration of counterpart and subsequent scratching on sample surface indicating abrasive wear. During sliding the sample experiences deformations that increases with number of cycles and after a critical accumulation, the nucleated part is removed from the surface. The removal materials from the surface of the sample in the form of debris and flakes (magnified SEM images in Fig. 5) are an indication of adhesive wear. While with increasing number of cycles the wear track gradually widens in all cases, the wear trace of $ZrO_2$ coated samples was narrower compared to S-P. This is once again attributable to the higher hardness of $ZrO_2$ which restricted penetration of counterpart. The presence of O in the EDS analysis indicated the occurrence of oxidative wear on the sample. The higher oxygen content in S-P sample is indicative of the higher temperature reached by the contact region due to larger friction experienced which promotes the oxidation reaction. Thus the mechanism of here can be considered abrasive, adhesive along with oxidation. This is in line with the findings of some other researchers where $ZrO_2$ coating improved the wear properties of the Ti6Al4V [31, 32]. Yuan *et al.* combined laser induced dimple pattern followed by zirconization of Ti6Al4V with double glow plasma technique to achieve 26.6% reduction in wear rate [31]. Berni *et al.* could observe 18% reduction in the wear rate in pulsed plasma deposited zirconia coating on Ti6Al4V [32]. In comparison to these, the wear rate reduced by 25%-50% for S-200 in comparison to pristine sample in present case. All these finding shows, PLD is an excellent choice for protective coating of biomaterials where the results for coating done at higher substrate temperature are superior. Further analysis on effect of other substrate temperature (both lower and higher) is however required to fully understand the properties $ZrO_2$ coating to improve performance of bioimplants at a long distance addressing orthopaedic application and durability.

*4.2 Antibacterial test*

Bacterial adhesion during or after implantation can lead to biofilm formation, a complex process influenced mainly by the bacterial properties and material surface characteristics such as chemical composition, surface charge, wettability, and physical configuration [33-34]. $ZrO_2$ is a biocompatible material with known antibacterial properties. Hence, PLD of $ZrO_2$ on Ti6Al4V is also expected to yield improved anti-bacterial performance which has been substantiated by our experimental observations.

The strength of bacterial adhesion depends on whether the bacteria donates or accepts electrons from the substrate surface [35]. $ZrO_2$, with a superior electron donor capability, can repel the bacteria leading to lower bacterial adhesion [36-38]. The adhesion, density and overall dose of two chosen bacteria, one gram positive (*S. aureus*) and one gram negative (*K. pneumonia*) is summarized in Figs.7 and 8. The total viable count analyzed after 4hr incubation indicated the antimicrobial effect of S-RT and S-200 against both the bacteria to be ~50% more in comparison to S- P. As can be seen from table-I, the reduced interfacial surface energy of the Ti6Al4V sample following $ZrO_2$ coating results in inferior adherence of bacteria [24, 39-40]. In live/dead bacteria analysis, there was a qualitative increase in the number of bacteria with compromised membranes on S-RT and S-200 for both *S. aureus* (Fig. 8a) and *K. pneumoniae* (Fig. 8b) as revealed by the increased red fluorescence as compared to S-P. The reduced bacterial colony counts on S-200 compared to S-RT can understandably be accounted by its low surface free energy. Thus there is a definite agreement in the observed cytological data and the corresponding physiological data of the viable count, which confirms the improved bactericidal properties of the $ZrO_2$ coated sample.

5. **Conclusion**

In this communication, the potential benefits of PLD of $ZrO_2$ thin film on Ti6Al4V sample in terms of superior wear resistance and enhanced bactericidal properties simultaneously in comparison to pristine sample have been addressed. Reduction in surface roughness with better adhesion and lower wettability was observed for the sample coated at 200 C substrate temperature. The coated sample showed significant improvement in wear resistance for all load conditions of 2N, 5N, and 7N against SS ball as counterpart. The sample showed about 30% reduction in CoF and 50% reduction in wear rate, indicating superior conservational properties of $ZrO_2$ coating by PLD against wear. The wear analysis using SEM and EDS indicated occurrences of abrasive, adhesive and oxidation wear on all the samples. Reduction in the surface free energy and superior electron donor capability of $ZrO_2$ restricted adhesion and biofilm formation of *S. aureus* and *K. pneumonia* bacteria indicated improved antibacterial behaviour of Ti6Al4V post coating. Hence, PLD of $ZrO_2$ coating could improve mechanical, tribological and antibacterial properties of widely used Ti6Al4V bio-alloy simultaneously.

**Acknowledgment**

Authors acknowledge Dr. T. S. R. C. Murthy, MP&CED and Dr. A. K. Sahu, G&AMD, BARC for fruitful discussions.

Figures:

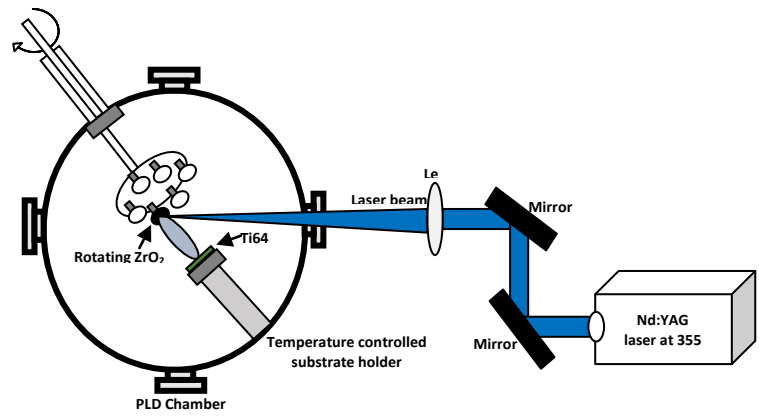

**Fig. 1:** *Experimental illustration of PLD of ZrO$_2$*

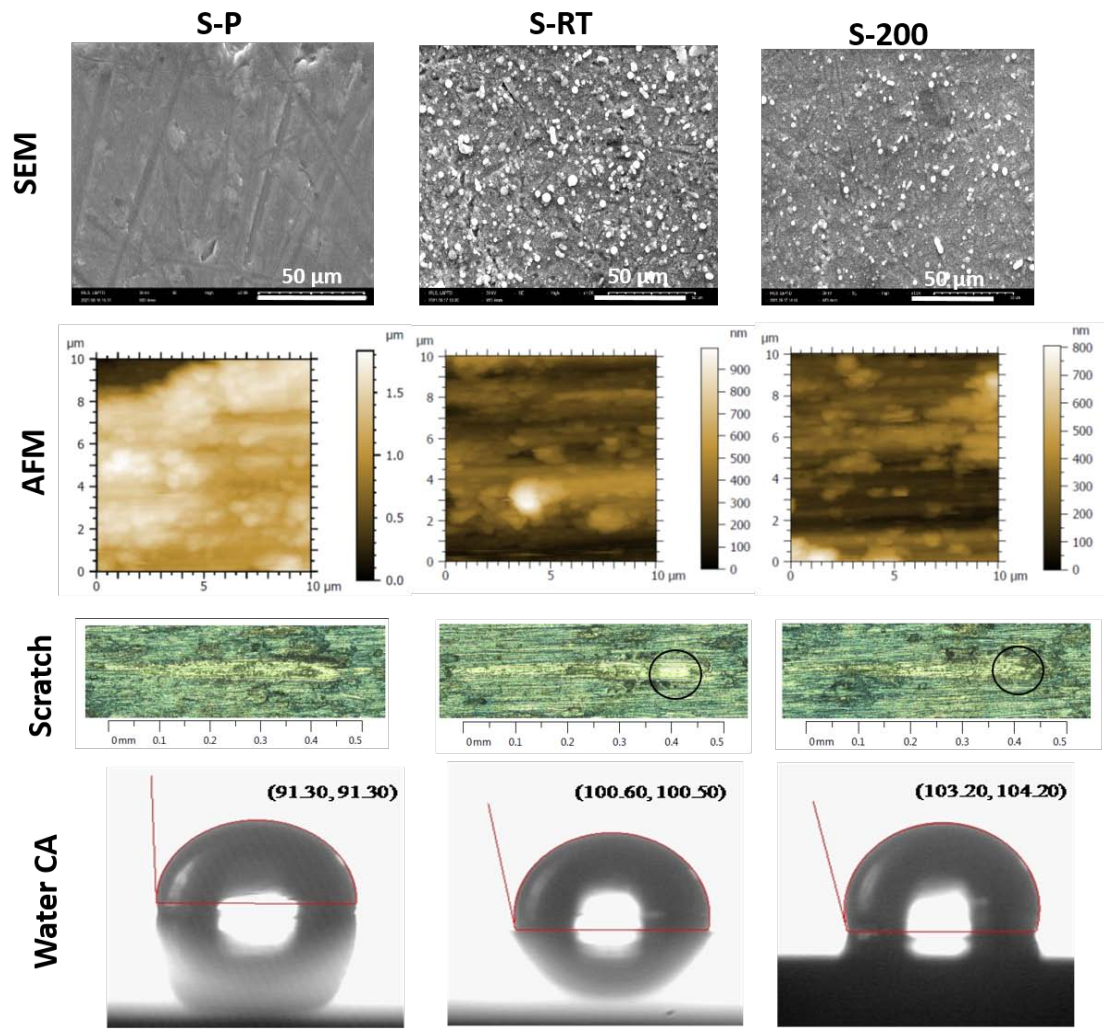

*Fig. 2: SEM image, AFM image, optical image of scratched line and water contact angle of S-P, S-RT and S-200 samples*

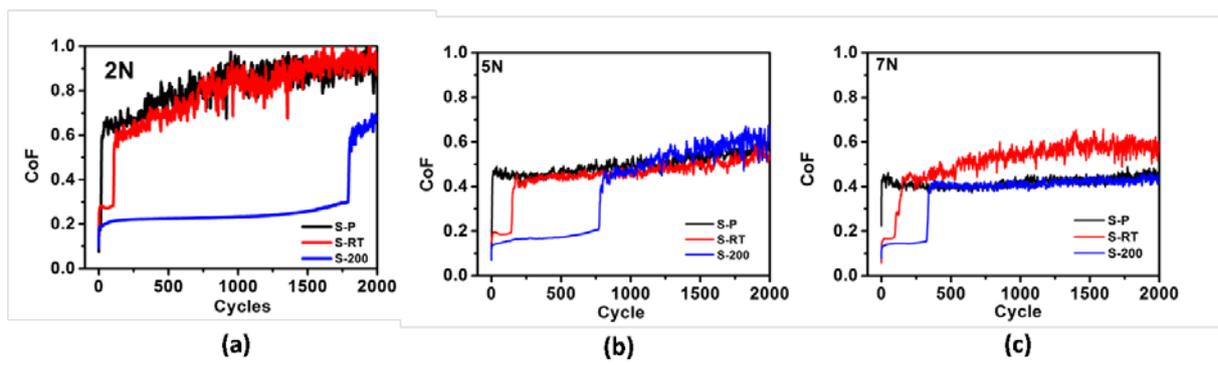

**Fig. 3:** *CoF of S-P, S-RT and S-200 samples at (**a**) 2N, (**b**) 5N and (**c**) 7N*

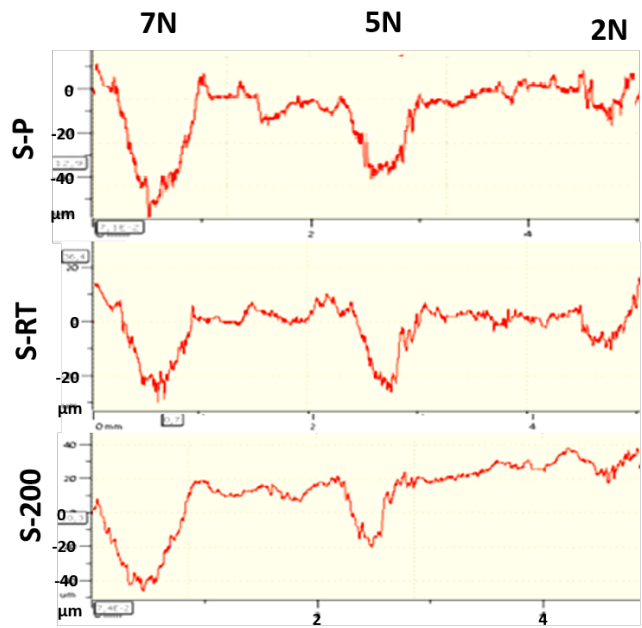 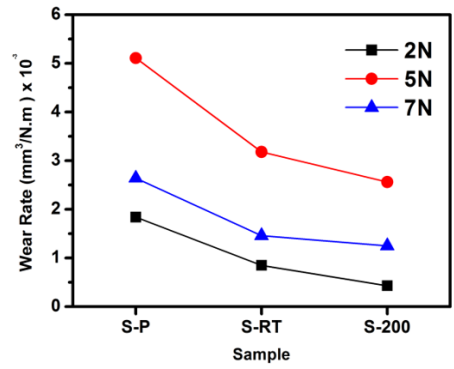

**Fig 4:** (**a**) *Two-dimensional depth profiles of wear trace* (**b**) *Wear rates of S-P, S-RT and S-200 samples as a function of load*

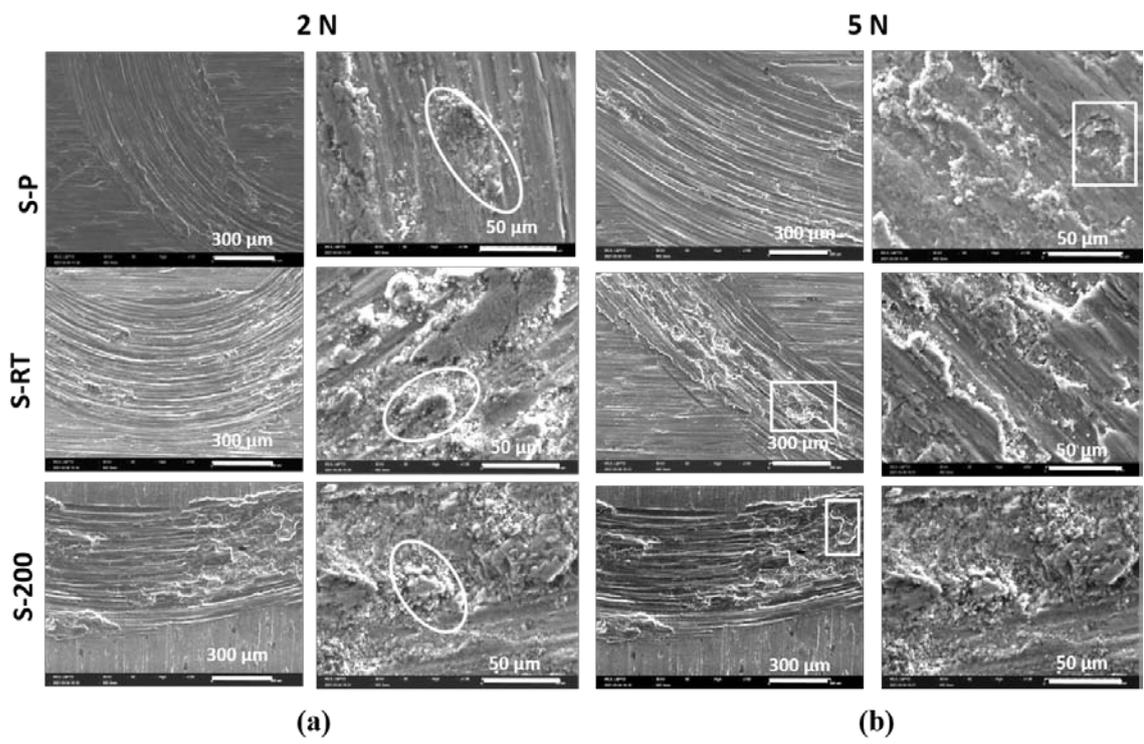

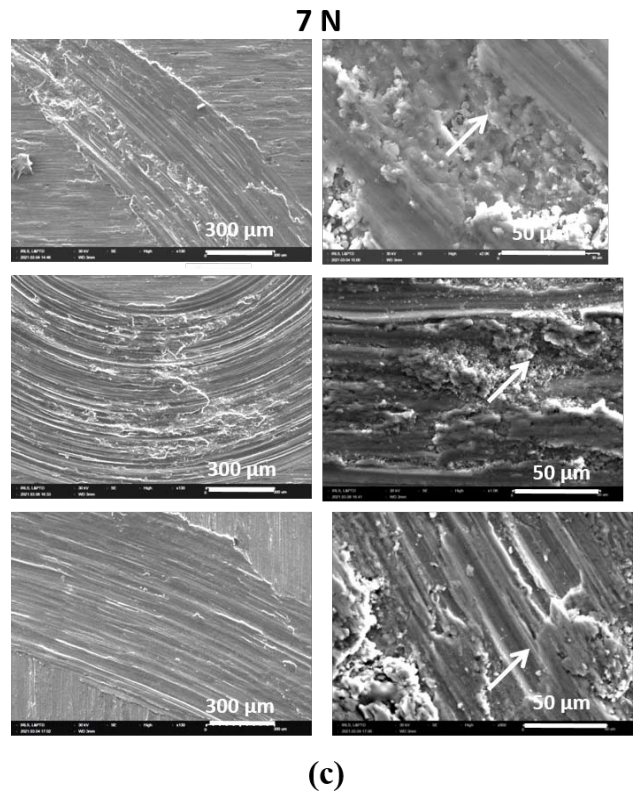

***Fig. 5:*** *SEM image of wear track of S-P, S-RT and S-200 samples at (a) 2N, (b) 5N and (c) 7N*

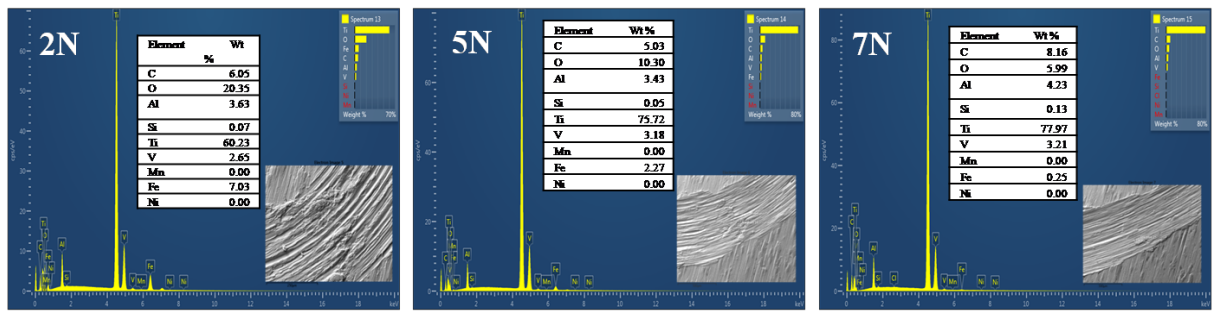

**Fig. 6:** *EDS analysis on wear track of S-200 at load of (**a**) 2N (**b**) 5N and (**c**) 7N*

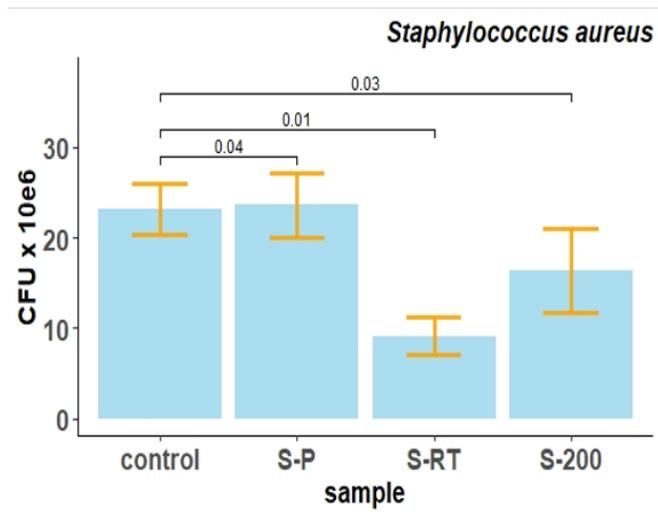 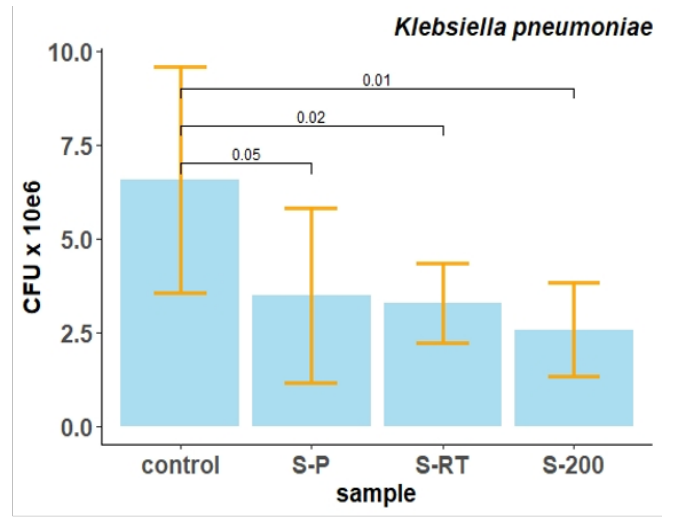

**Fig. 7:** *Number of cells in control, S-P, S-RT and S-200 (a) S. aureus and (b) K. Pneumonia*

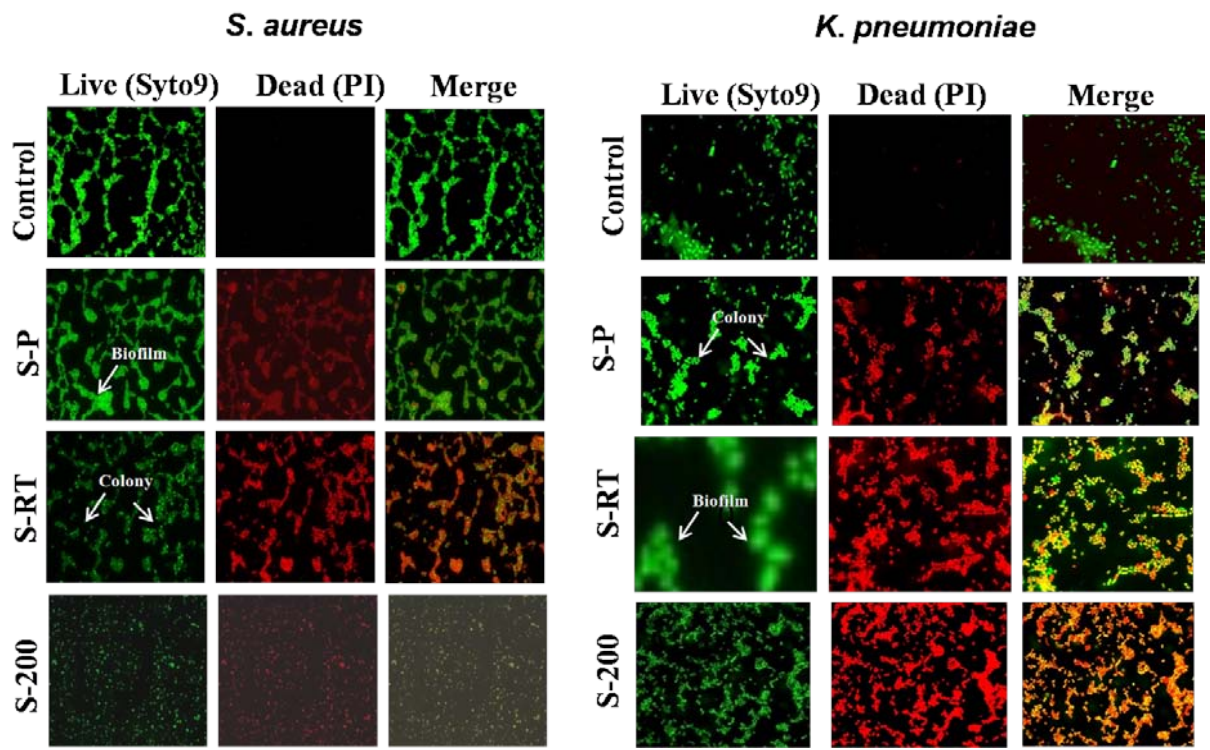

*Fig. 8: Fluorescence image of (a) S. aureus and (b) K. Pneumonia on control, and S-P, S-RT, S-200 samples*